\documentclass[10pt, a4paper]{article}

\usepackage[final]{lrec-coling2024} % this is the new style

\usepackage{xcolor}
\usepackage{hyperref}
 \definecolor{darkblue}{rgb}{0, 0, 0.5}
  \hypersetup{colorlinks=true, citecolor=darkblue, linkcolor=darkblue, urlcolor=darkblue}

\usepackage{xstring}
\usepackage{ulem}
\usepackage{amssymb}
\usepackage{multirow}
\usepackage{color}
\usepackage{booktabs}
\usepackage{arydshln}
\usepackage{stfloats}
\usepackage{subfigure}
\usepackage{caption}
\usepackage{amsmath}
\usepackage{ulem}

\makeatletter 
  \newcommand\figcaption{\def\@captype{figure}\caption} 
  \newcommand\tabcaption{\def\@captype{table}\caption} 
\makeatother

\newcommand{\dsname}{CitationR}
\newcommand{\taskname}{RMC}
\newcommand{\framename}{\taskname Net}

\title{Recommending Missed Citations Identified by Reviewers: A New Task, Dataset and Baselines}

\name{Kehan Long$^{1}$, Shasha Li$^{1,2,\dagger}$ \thanks{$\dagger$ Corresponding authors.}, Pancheng Wang$^{1}$, Chenlong Bao$^{1}$, \\ {\bf \large  Jintao Tang$^{1,\dagger}$, Ting Wang$^{1,\dagger}$ } }

\address{$^{1}$School of Computer, National University of Defense Technology \\
         $^{2}$Key Laboratory of Software Engineering for Complex Systems, National University of Defense Technology \\
         lishasha198211@163.com \\
         \{longkehan15, wangpancheng13, baochenglong, tangjintao, tingwang\}@nudt.edu.cn\\}

\abstract{
Citing comprehensively and appropriately has become a challenging task with the explosive growth of scientific publications. 
Current citation recommendation systems aim to recommend a list of scientific papers for a given text context or a draft paper. 
However, none of the existing work focuses on already included citations of full papers, which are imperfect and still have much room for improvement. 
In the scenario of peer reviewing, it is a common phenomenon that submissions are identified as missing vital citations by reviewers.
This may lead to a negative impact on the credibility and validity of the research presented. 
To help improve citations of full papers, we first define a novel task of \underline{R}ecommending \underline{M}issed \underline{C}itations Identified by Reviewers (\taskname) and construct a corresponding expert-labeled dataset called \dsname.
We conduct an extensive evaluation of several state-of-the-art methods on \dsname.
Furthermore, we propose a new framework \framename \ with an \textit{Attentive Reference Encoder} module mining the relevance between papers, already-made citations, and missed citations.
Empirical results prove that \taskname \ is challenging, with the proposed architecture outperforming previous methods in all metrics.
We release our dataset and benchmark models to motivate future research on this challenging new task.
 \\ \newline  
}

\begin{document}

\maketitleabstract

\section{Introduction}

Citations are essential in many writing scenarios, especially in academic writing \cite{review2020}.
Academic researchers constantly refer to credible literature in their research fields to support their arguments and provide readers with a plausible explanation of the content of their manuscripts.
Those reliable sources should be correctly cited, and this is where the concept and significance of citations enter the picture \cite{bookcite}. 
A broad and critical literature survey is an essential component of any scientific research, as it provides a foundation for developing research questions, designing experiments, and interpreting results \cite{booklr}.
However, the exponential growth of scientific publications has made it increasingly challenging for researchers to conduct thorough literature reviews and make comprehensive and appropriate citations.

The task of citation recommendation (CR) has been introduced by \cite{He2010ContextawareCR}, aiming to automatically recommend appropriate citations for a given text context or a draft paper.
Here we denote a scientific paper with only content as a \textit{draft paper} or \textit{manuscript} and a paper containing both content and complete citations as a \textit{submission paper} or \textit{full paper}. 
Specifically, researchers collect published scientific papers and take already cited papers in the reference section as labels for training models to predict.
% \cite{DACR, Huang2015ANP, Ebesu2017NeuralCN, acl18, lcr, citeomatic} generate recommendations by learning semantic representations of the content of input draft papers using neural networks.
\begin{figure}[t]
    \centering
    \includegraphics[width=0.95\linewidth]{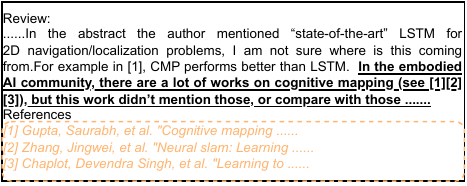}
    \caption{An example of missed citations identified by reviewers extracted from \url{https://openreview.net/forum?id=R612wi\_C-7w}. Italic and colored texts represent papers mentioned in reviews, among which, enclosed by the dashed border, are those reviewers recommend citing. Underlined and bolded texts indicate reasons why those citations are missed and necessary.}
    \label{fig:reviews}
\end{figure}{}
Some works \cite{Huang2015ANP, citeomatic, hybrid20} rely on semantic representations of the content of draft papers learnt by neural networks to generate recommendations.
% \cite{Ren2014ClusCiteEC, jiang_cross-language_2018, Ali2021CitationRE, Pornprasit2022EnhancingCR, Wang2022CollaborativeFW, Xie2021GraphNC, chen_citation_2019} further adopt embeddings learnt from graphs to recommend citations. 
Some works \cite{Ren2014ClusCiteEC, Xie2021GraphNC, Wang2022CollaborativeFW} further adopt embeddings learnt from graphs to recommend citations. 
Although some of those studies construct a citation network based on partial citations of input papers, they equally treat these citations and are unaware of critical citations that significantly impact the research's foundation.
In all, previous studies mainly focus on manuscripts or give equal weights to all citations, neglecting the potential imperfections of already included citations and the negative impact of missing vital citations on the research's comprehensiveness and innovativeness.

It is not a rare case that submissions are considered to lack vital citations by reviewers, as illustrated in Figure \ref{fig:reviews}.
The omission of these essential papers in the submission often leads to deficiencies in terms of credibility, comprehensiveness, and innovation. 
That is to say, citations recommended by reviewers may have a great influence on the research foundation of submissions.
Inspired by this common phenomenon in the peer review process, we formulate (\S \ref{sec:task}) and study a novel task of \underline{R}ecommending \underline{M}issed \underline{C}itations Identified by Reviewers (\taskname).
Previous CR tasks generate recommendations by taking already cited papers provided by authors as labels for training. 
RMC enhances the reference sections of submissions by considering citations recommended by reviewers as golden labels.
Considering the data flow, CR is similar to \taskname. However, they differ in several aspects:
\begin{itemize}
    \item \textbf{Recommend for}: Local CR recommends citations for a text context where specific citations should be made. Global CR takes a draft paper with no or partial citations as input. \taskname \ identifies missed citations of a full paper with complete citations provided by the author.
    \item \textbf{Guarantee}: The golden citations of both CR tasks are obtained from the reference sections in papers, which are written and guaranteed by corresponding authors. In \taskname, golden citations are identified by experienced experts from top-tier conferences.
    \item \textbf{Relevance}: For local CR, golden citations are highly related to input texts since that is where those citations should be made. Global CR aims to recommend a whole reference list, with some citations highly related and some less important and even replaceable. For \taskname, all golden citations are considered highly related and important by reviewers.
\end{itemize}

Additionally, we curate a novel high-quality dataset, \dsname, by extracting recommended citations in reviews from NeurIPS and ICLR (\S \ref{sec:dataset}).
In total, we collect 76,143 official reviews and 21,598 submissions, among which around 35\% of submissions are identified as lacking citations.
Moreover, to better replicate the actual situation in which researchers search for papers to cite, we establish a larger and more challenging version of  \dsname. This version includes additional 40,810 papers published in top venues that reviewers frequently recommend citations from.

We adapt and evaluate a wide range of existing state-of-the-art methods on our dataset and task formation, including four groups (\S \ref{sec:baselines}): (1) traditional sparse retrieve models, (2) traditional citation recommendation models, (3) pre-trained scientific document encoding models, and (4) large language models.
We further propose a novel framework \framename \ based on an \textit{Attentive Reference Encoder (ARE)} module and contrastive learning objectives to solve this task. 
One key challenge in \taskname \  is to assess the correlation between missed citations, already included citations and the content of submissions.
Our designed \textit{ARE} aims to effectively fuse both the content and reference sections of submissions.
We use the ``Citation-Informed Transformer" \cite{scincl} as the text encoder in our framework, which can be readily adapted to other encoders.

In experiments, we show that our method outperforms all previous methods in all metrics on \dsname. 
Ablation and parameter studies further prove the effectiveness of our approach.
However, compared to traditional CR datasets, the performances of all methods on \dsname \ are much worse, highlighting the complexity of \dsname.

To conclude, our contributions are threefold:
\begin{itemize}
    \item We introduce a new challenging task of recommending missed citations identified by reviewers, which is built from a common phenomenon in the peer review process and aims to avoid the reliability and novelty of research being undermined due to missing vital citations.
    \item We develop a novel high-quality dataset containing submission-citation pairs extracted from real reviews, which are actually labeled by experienced experts from top-tier conferences and are easy to extend with more reviews coming out annually.
    \item We evaluate several mainstream methods in other similar research tasks on our proposed \dsname \ and establish a new method. Our proposed method achieves the best and can serve as a solid baseline for future research. All data and code are publicly available\footnote{\url{https://github.com/ChainsawM/RMC}}.
    % \item We establish a new method and conduct extensive experiments on two versions of \dsname \  dataset. Our proposed method performs best and can serve as a solid baseline for future research\footnote{All code and data would be available.}.
\end{itemize}

\section{Related Work}

We introduce the related work from three aspects, including citation recommendation (CR), transformer-based scientific document encoders, and other review-mining tasks.

\subsection{Citation Recommendation}
\label{sec:citation_recommendation}

% Based on the degree the input text covers the source paper, citation recommendation can usually be divided into two types: global citation recommendation \cite{gupta_scientific_2017, jiang_cross-language_2018, citeomatic, cai_generative_2018, chen_citation_2019, Hu2020FusionOD}, which recommends citations for a draft paper, and local citation recommendation \cite{Tang2009ADA, He2010ContextawareCR, He2011CitationRW, Huang2012RecommendingCT, Huang2015ANP, Ebesu2017NeuralCN, Yin2017PersonalizedCR, acl18, DACR, lcr}, which recommends citations for a short text context. 
Based on the degree the input text covers the source paper, citation recommendation (CR) can usually be divided into two types \cite{He2010ContextawareCR}: global citation recommendation \cite{gupta_scientific_2017, jiang_cross-language_2018, Hu2020FusionOD}, which recommends citations for a draft paper, and local citation recommendation \cite{Ebesu2017NeuralCN, Yin2017PersonalizedCR}, which recommends citations for a short text context. 
Although early works \cite{Tang2009ADA, He2011CitationRW, Huang2012RecommendingCT} use probability distributions to represent and analyze the relevance of documents, the advancement and prevalence of neural networks have made embedding-based models \cite{Huang2015ANP, cai_generative_2018, acl18, DACR} the mainstream approach.
Recently, \citet{citeomatic} use shallow feed-forward networks to learn representations of content and metadata of papers and introduce a contrastive learning objective for training the model.
\citet{lcr} adopt hierarchical transformer layers as paper encoders and use the same contrastive learning objective  for training.

Besides text, citations are a vital factor in measuring the similarity of scientific papers. 
Node representations learned from the citation graph can be taken as representations of papers \cite{cikm14, Pornprasit2022EnhancingCR}. 
However, constructing a citation graph requires a huge number of papers with dense citations, which is beyond our current collected dataset. Thus, graph-based methods are not considered in this paper.

\subsection{Transformer-based Scientific Document Encoders}
Pre-trained Language Models (LMs) \cite{bert} based on transformer architecture have shown their surprising ability on numerous natural language  processing tasks.
Adapting to scientific domain corpora \cite{scibert} further improved the performance of LMs and dominates various scientific document processing tasks, such as explaining relationships between scientific papers \cite{Luu2020ExplainingRB}, scientific fact-checking \cite{Cai2022COVIDSumAL}, citation recommendation \cite{lcr, Medic2022LargescaleEO}, and so on.

Recently, several works try to leverage citations between scientific documents to enhance their representation learning when pre-training LMs. 
SPECTER \cite{specter} leverages citations as a signal for document-relatedness and formulates this into a triplet-loss contrastive learning objective. 
% The representations of SPECTER are effective across a variety of downstream tasks without task-specific fine-tuning.
CiteBERT \cite{citebert} trains SciBERT with the task of cite-worthiness detection and LinkBERT \cite{linkbert} fine-tunes BERT on the extra task of document relation prediction.
SciNCL \cite{scincl} uses citation graph embeddings for a more informative selection of negative examples with the same contrastive learning objective as SPECTER.
Transformer-based scientific document encoders with citation information introduced generally achieve better performance on document-level tasks.
We adapt and evaluate their performance on \taskname \  and apply them as the text encoder in our proposed framework to help solve the task.

Besides, Large Language Models (LLMs), which undergo extensive pretraining on diverse textual sources, have showed remarkable capabilities in text generation, language understanding, and context preservation \cite{Zhou2023ACS, Zhu2023LargeLM}. 
LLMs have been applied across various research fields, such as natural language processing (NLP) \cite{gpt3, llama}, code generation \cite{codex, codegeex}, and recommender systems \cite{Fan2023RecommenderSI, Liu2023ONCEBC, Wu2023ASO, Hou2023LargeLM}.
In this paper, we include LLMs as baselines and design several prompts to evaluate their performance on \taskname.

\subsection{Other Review-Mining Tasks}
In addition to identifying missed citations in reviews, there are other review-related tasks that can be roughly categorized into three main areas:
(1) Using the content of reviews to predict the citation count of submitted papers \cite{re_li_19, re_li_22}, predict final decisions \cite{re_wang_18, re_deng_20, re_kumar_22}, and predict aspect scores \cite{re_deng_20, re_li_20}.
(2) Analyzing the content of reviews for argument mining \cite{re_hua_19}, sentiment analysis \cite{re_wang_18, re_chak_20}, and grading reviews \cite{re_arous_21, Bharti2022AMF}.
(3) Investigating the writing patterns of reviewers and exploring ways to automate the peer review process \cite{re_wang_20, re_yuan_21, re_lin_21}.
Different from \taskname , these tasks primarily involve utilizing the content of reviews for various purposes.

\section{Task Formulation}\label{sec:task}
% Before describing our dataset and model, we first formulate the task of recommending missed citations identified by reviewers and introduce several basic concepts.
Let $S=\{s_1, \dots ,s_m\}$ be the set of $m$ submissions. 
From their reviews, a set of missed citations recommended by reviewers $R = \{ r_{1,1}, \dots, r_{1,p}, \dots, r_{m,1}, \dots, r_{m,q} \}$ can be extracted, where $r_{m,q}$ is the $q$-th recommended paper for submission paper $s_m$. 
Besides, to better mimic the real situation, a set of $n$ candidate papers $C=\{ c_1, \dots, c_n \}$ is collected based on the characteristics of $R$ (\S \ref{candidataextension}).

For a recommendation model, given the set of all papers $P = S \cup R \cup C$, its input is a submission paper $s$, and it is supposed to calculate a rank score for each candidate paper in $P - \{s\}$ and output a paper list according to the descending rank scores.

In the task of \taskname , we define $p = \{T,A,T_{r_1},\dots ,T_{r_n}\}$ represent a paper from the total paper collection $P$, where $w \in W$ is a word from the vocabulary set $W$, $T = [w_{1}^t, \dots, w_{i}^t]$ is the title consists of $i$ words, $A = [w_{1}^a, \dots, w_{j}^a]$
is the abstract consists of $j$ words and $T_{r_n} = [w_{1}^{r_n}, \dots, w_{m_{n}}^{r_n}]$ is the title of the $n$-th paper from the reference of $p$.

\section{Dataset Construction}\label{sec:dataset}

\begin{figure*}[t] 
\centering
    \begin{minipage}[t]{0.48\textwidth}
    \centering
    \includegraphics[width=\linewidth, trim=20cm 20cm 20cm 26.5cm, clip]{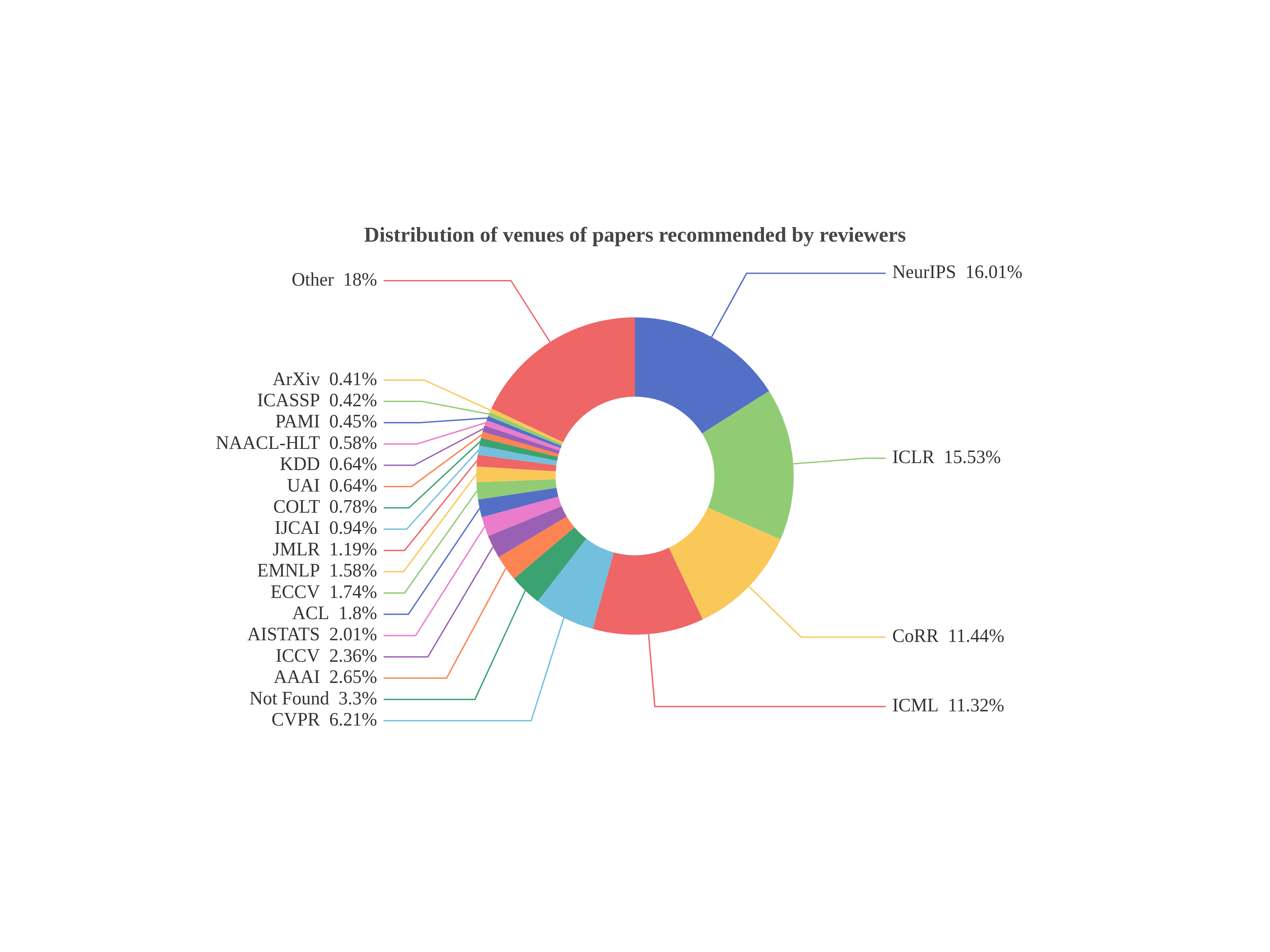} 
    \figcaption{Distribution of venues of extracted citations recommended by reviewers.} 
    \label{fig:venues} 
    \end{minipage} 
    \hfill
    \begin{minipage}[t]{0.48\textwidth}
    \centering
    \includegraphics[width=\linewidth]{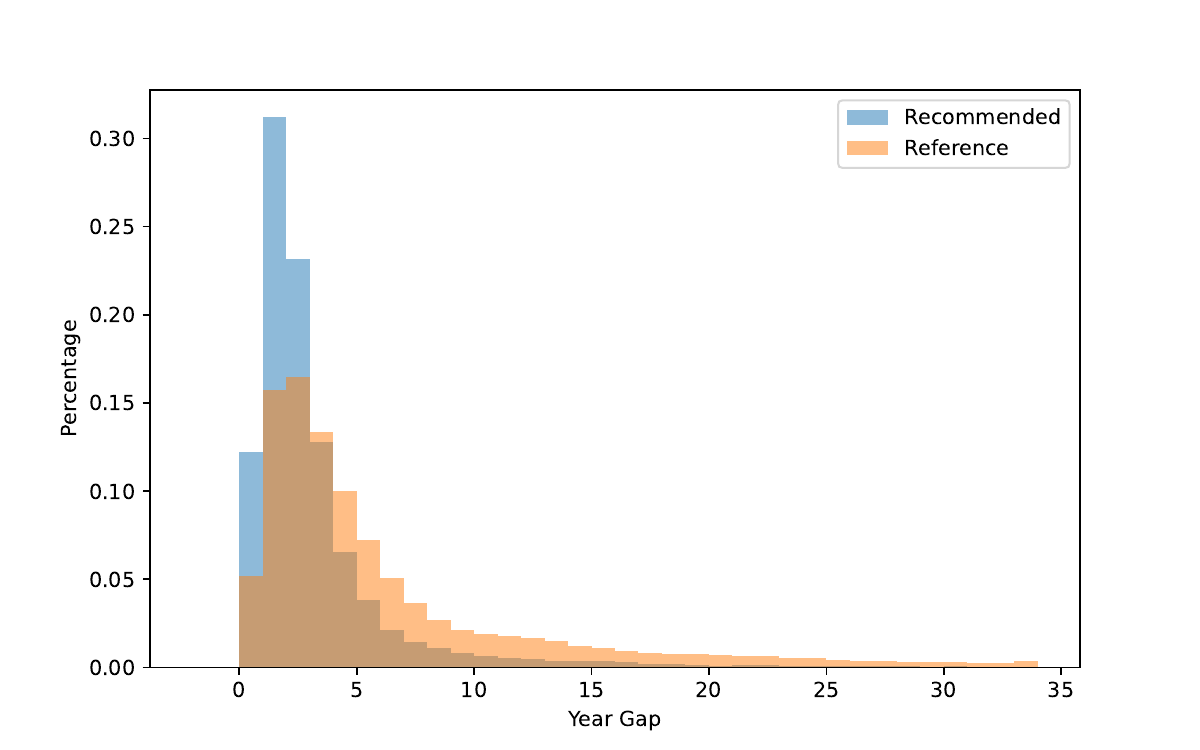} 
    \figcaption{Distribution of year gaps between submissions and their citations.} 
    \label{fig:year_gap} 
    \end{minipage}
\end{figure*}

In this section, we present the building process and details of two versions of \dsname \ dataset.

\subsection{Review Collection}
The first step is review collection.
We download reviews of scientific submissions from two conference sources: the NeurIPS\footnote{\url{https://proceedings.neurips.cc/}} and the ICLR\footnote{\url{https://openreview.net/group?id=ICLR.cc/}}. 
From the former source, we collect all accepted papers and corresponding reviews for NeurIPS 2013-2021, a total of 34,613 reviews for 9,526 papers. 
From OpenReview, we collect all submissions to ICLR 2017-2022, a total of 41,530 official anonymous reviews for 12,072 papers.

\subsection{Citation Extraction}
Peer review has been adopted by most journals and conferences to identify important and relevant research.
Although various guidelines\footnote{\url{https://icml.cc/Conferences/2023/ReviewerTutorial}} or tutorials\footnote{\url{https://neurips.cc/Conferences/2022/ReviewerGuidelines}} about how to write good reviews have been proposed, there are no unified standards on the format of reviews, let alone how to cite external resources.

Empirically, we classify all mentions of papers into two categories. The first type pertains to papers that have already been cited by the authors but are mentioned by reviewers in their critiques. 
These mentions typically appear as brief phrases, such as a concatenation of the author name and publication year of the paper, or short phrases enclosed in brackets, such as ``[ref X]'', and so on.
The second type of mentions includes papers that the authors do not cite but that are identified and recommended by reviewers to be included.
Typically, these papers are mentioned through a formal reference section attached to the end or via URL links to external resources.
Considering simplicity and practice, we make the intuitive assumption that missed citations in reviews are mentioned in the format of reference strings or URL links. 
Then we extract these mentions using regular expressions and manually corrected samples that are found invalid in the latter steps.

\subsection{Paper Alignment}
For papers mentioned in the format of URL links, we directly download according files and remove those that are not scientific papers. 
For reference strings, we try to align them to scientific papers via searching in bibliographic databases like DBLP\footnote{\url{https://dblp.uni-trier.de/}} and Semantic Scholar\footnote{\url{https://www.semanticscholar.org/}}. 
In general, we adopt three rules to judge the alignment of reference strings and scientific papers: (1) The title of the retrieved paper appears in the reference string, (2) titles of retrieved papers belonging to the same reference string from two sources are matched, (3) and unmatched samples are manually checked. 
For downloaded papers, the tool Doc2json\footnote{\url{https://github.com/allenai/s2orc-doc2json}} is used to extract their metadata and bibliographies.

In all, we collect 14,520 unique papers recommended by reviewers. Out of 21,598 collected submissions, 7,528 papers (around 35\%) are identified as missing citations. 
Out of 76,143 collected reviews, 9,100 (around 12\%) reviews contain citations recommended by reviewers. 
The average number of recommended citations per submission paper is around 2.5. 
Apparently, it is not a rare case that reviewers regard submissions as lacking important citations and recommend papers to cite.

\begin{table*}[t]
\centering
% \resizebox{0.95\textwidth}{!}{
\begin{tabular}{lcccccc}
\hline
\multirow{2}{*}{\textbf{Section}} & \multicolumn{3}{c}{\textbf{\# Samples}} & \multirow{2}{*}{\begin{tabular}[c]{@{}c@{}}\textbf{\# Submissions}\\ (valid / collected)\end{tabular}} & \multirow{2}{*}{\begin{tabular}[c]{@{}c@{}}\textbf{\# Reviews}\\ (valid / collected)\end{tabular}} & \multirow{2}{*}{\textbf{Publication years}} \\ \cline{2-4}
                         & train     & val     & test    &                                                                                              &                                                                                          &                                    \\ \hline
ICLR                     & 10,646    & 1,556    & 1,492    & 5,127 / 12,072                                                                               & 5,872 / 41,530                                                                           & 2017-2022                          \\
NeurIPS                  & 4,257     & 566     & 582     & 2,401 / 9,526                                                                                & 3,228 / 34,613                                                                           & 2013-2021                          \\ \hline
total                    & 14,903    & 2,122   & 2,074   & 7,528 / 21,598                                                                               & 9,100 / 76,143                                                                           & 2013-2022                          \\ \hline
extended                 &           &         &         & 0 / +40,810                                                                                  &                                                                                          & 2009-2022                          \\ \hline
\end{tabular}
% }
\caption{Statistics of \dsname \ dataset. Here "valid" means missed citations are found from those reviews and for those submissions. }
\label{tab:statistics_detail}
\end{table*}

\subsection{Candidate Extension}\label{candidataextension}
To better mimic the real situation where researchers search for papers to cite, we also collect a set of candidate papers that featured the same as papers recommended by reviewers in some aspects. 
As illustrated in Figure \ref{fig:venues}, reviewers in NeurIPS and ICLR mostly recommend papers from top-tier conferences, and NeurIPS and ICLR are just the two venues reviewers most frequently recommend papers from. 
Moreover, as illustrated in Figure \ref{fig:year_gap}, compared to papers already cited by authors (in the reference section), papers recommended by reviewers are more up-to-date. 
Among papers we extracted, those published no more than three years earlier than submissions count more than 60\%, and that of papers in the reference sections is no more than 40\%. 

Thus, considering availability and the above characteristics, we collect 40,810 papers published no more than three years earlier than recommended papers from publicly available top-tier conferences, including AAAI, ACL, AISTATS, COLT, CVPR, EMNLP, ICCV, ICML that range from 2009 to 2022. By adding collected candidate papers, we get the extended version of \dsname.

Finally, we split the dataset into training, validation, and test sets roughly in the ratio 8:1:1 based on publication years.
% Following previous works, papers published in earlier years are used for training, and the newest papers are randomly selected to be included in the validation and test sets. 
Detailed statistics of the split dataset are listed in Table \ref{tab:statistics_detail}.

\begin{figure*}[ht]
    \centering
    \includegraphics[width=0.9\textwidth]{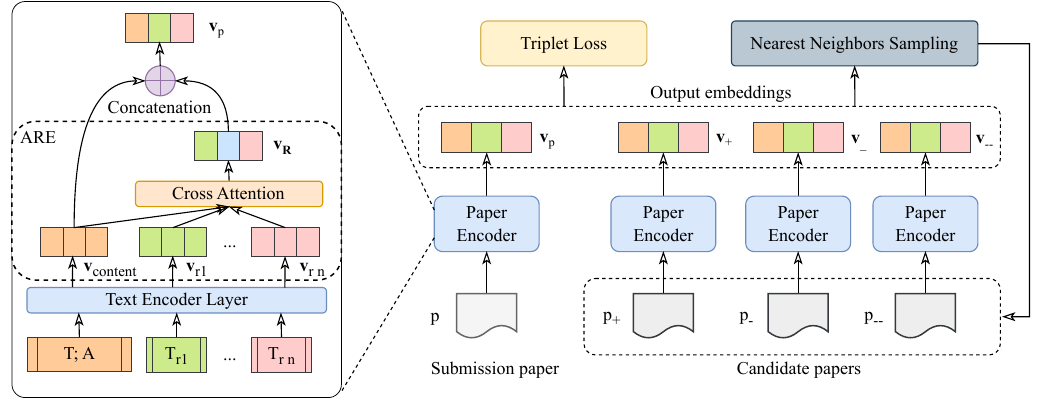}
    \caption{The overall architecture of \framename, which consists of three parts: (1) \textit{Paper encoder} (left) generates representations of papers with an \textit{Attentive Reference Encoder (ARE)} part mining the reference sections. (2) \textit{Triplet Loss} (upper middle) computes the loss fusing positive samples and negative samples of different levels. (3) \textit{Nearest Neighbors Sampling} (upper right) obtains negative samples of different levels based on output embeddings and their textual similarities to the submission paper.}
    \label{fig:model}
\end{figure*}{}

\section{Methodology}

% In this section, we first formulate some preliminaries. Then we introduce our proposed method in detail, whose overall framework is shown in Figure \ref{fig:model}.
In this section, we introduce the \framename \ in detail, whose overall framework is shown in Figure \ref{fig:model}.

\subsection{Paper Encoder}
% Paper encoder is one of the core components in our model that aims to learn the embeddings of papers from their texts. 
Paper encoder aims to learn the embeddings of papers from their texts. 
Transformer-based encoder is adopted as the text encoder in our proposed model and can be easily replaced by other models.
For a paper $p$, the concatenation of its title and abstract with an additional separator token inserted between them is fed into BERT, and a series of hidden states can be obtained:
\begin{align}
    \textbf{h}_{[{\rm C}]}, \textbf{h}_{1}^t, \dots,  \textbf{h}_{j}^a = {\rm BERT}(&[{\rm C}], w_{1}^t, \dots, w_{i}^t, \notag \\
    &[{\rm S}], w_{1}^a, \dots, w_{j}^a)
\end{align}
% \begin{equation}
%     \textbf{h}_{[{\rm C}]}, \textbf{h}_{1}^t, \dots,  \textbf{h}_{j}^a = {\rm BERT}([{\rm C}], w_{1}^t, \dots, w_{i}^t, [{\rm S}], w_{1}^a, \dots, w_{j}^a)
% \end{equation}{}
where $[{\rm C}]$ denotes the special $[{\rm CLS}]$ token in BERT that is added to the front of a sequence, and $[{\rm S}]$ is the sentence separator token $[{\rm SEP}]$. 
Following \cite{bert}, the hidden state of $[{\rm C}]$ is used as the representation of the content of the input paper:
\begin{equation}
     \textbf{v}_{\rm content} = \textbf{h}_{[{\rm C}]}
\end{equation}{}
For an already cited paper in the reference section of $p$, its title $T_{r_x} \in P$ is input into BERT:
\begin{equation}
        \textbf{h}_{[{\rm C}]}^{r_x}, \textbf{h}_{1}^{r_x}, \dots,  \textbf{h}_{m_{x}}^{r_x} = {\rm BERT}([{\rm C}], w_{1}^{r_x}, \dots, w_{m_{x}}^{r_x}, [{\rm S}])
\end{equation}{}

Similarly, we obtain the representation of an already cited paper in the reference section:
\begin{equation}
     \textbf{v}_{r_x} = \textbf{h}_{[C]}^{r_x}
\end{equation}{}
Intuitively, a paper's citing pattern can be exploited from the relations between itself and already cited papers. 
To model existing citations, we feed all embeddings of already cited papers into an \textit{Attention} layer \cite{attention17} and calculate the embedding of the reference as follows:
\begin{equation}
\label{attention}
{\boldsymbol {\rm v}}_{R} = \sum\limits_{x=1}^{n} {\boldsymbol  {\rm w}}_x {\boldsymbol {\rm v}}_{r_x}
\end{equation}
\begin{equation}
\label{alpha}
{\boldsymbol  {\rm w}} = {\rm softmax}({[{\boldsymbol {\rm v}}_{r_1}, \dots ,{\boldsymbol {\rm v}}_{r_n}]}^{\top} \cdot \textbf{v}_{\rm content})
\end{equation}
where $\boldsymbol  {\rm w}$ is the weight vector and the scalar ${\boldsymbol  {\rm w}}_x$ is its $x$-th element.

Finally, in order to recommend missed citations relating to both the content and citation pattern, and avoid meaningless duplicates, we linearly combine the embeddings of content and reference to get the final representation of the input paper:
\begin{equation}
\label{combination}
{\boldsymbol {\rm v}}_{p} = (1 - \alpha) {\textbf{v}_{\rm content}} + \alpha {\boldsymbol {\rm v}}_{R}
\end{equation}
where $\alpha$ is the parameter to balance the content and existing citations.

\subsection{Triplet Loss}
In particular, each training instance at least contains a triplet of papers: a submission (query) paper $p$, a positive paper $p_{+}$, and a negative paper $p_{-}$. 
The positive paper is the recommended citation extracted from reviews, and the negative paper is a paper that is not cited by $p$ or recommended by reviewers.
Via previously introduced paper encoder, respective ${\boldsymbol {\rm v}}_{p}$, ${\boldsymbol {\rm v}}_{+}$, ${\boldsymbol {\rm v}}_{-}$ can be obtained, which represent the embeddings of papers from a training sample.
We then train the model using the following triplet margin loss function:
\begin{equation}
\label{loss}
\mathcal{L} = {\rm max}\{s({\boldsymbol {\rm v}}_{p}, {\boldsymbol {\rm v}}_{-}) - s({\boldsymbol {\rm v}}_{p}, {\boldsymbol {\rm v}}_{+}) + m, 0\}
\end{equation}
where $s$ is the similarity function and $m$ is the loss margin hyper-parameter sets the span over which the loss is sensitive to the similarity of negative pairs.
Following previous work \cite{citeomatic, lcr}, $s$ is defined as the cosine similarity between two document embeddings.

\subsection{Nearest Neighbors Sampling}
The definition of positive example papers  $p_{+}$ is straightforward, which are papers recommended by reviewers.
However, a careful choice of negative example papers may be critical for model performance.
We use two types of negative examples:
(1) \textbf{Random}: Randomly selecting a paper as a negative example typically results in easy negative examples.
(2) \textbf{Negative nearest neighbors}: Given a submission paper, we use the model's current checkpoint to obtain the top $K_n$ nearest candidates excluding golden papers.
The embeddings of these candidates have high similarities to the embedding of the input submission paper. 
Thus, we denote papers selected from those candidates as "hard negatives". 
It is expected that training the model to distinguish hard negative examples may improve overall performance. 
The checkpoint of the model is updated every $N_{iter}$ training iterations, at which point the nearest candidates are also updated.

\section{Experiments}

In this section, we evaluate our method and four groups of baselines on \dsname.
All reported scores are obtained by running models over at least three random seeds and are presented as percentage numbers with "\%" omitted.

\begin{table*}[t]
\renewcommand\arraystretch{1}
\centering
% \resizebox{0.95\textwidth}{!}{
\begin{tabular}{llcccccccc} 
\toprule[1pt]
\multicolumn{1}{l}{}    &   \multicolumn{1}{l}{Dataset$\rightarrow$}       & \multicolumn{4}{c}{\dsname}                                       & \multicolumn{4}{c}{Extended}                               \\ 
\cmidrule(l){3-6} \cmidrule(l){7-10}
\multicolumn{1}{l}{}    &   \multicolumn{1}{l}{Model$\downarrow$ / Metric$\rightarrow$} & MAP            & MRR           & NDCG         & R@10           & MAP            & MRR           & NDCG         & R@10            \\ 
\midrule
1   &   BM25                               & 10.79          & 17.47          & 26.36          & 19.41          & 7.11          & 12.22          & 22.00          & 13.76           \\
2   &   ChatGPT-turbo                        & 11.44          & 19.02          & 26.97          & 19.41          & 7.92          & 14.11          & 22.76          & 13.76           \\
3   &   LLaMA2-13b \shortcite{llama2}                      & 10.88          & 18.23          & 26.53          & 19.41          & 6.81          & 11.62          & 21.72          & 13.76           \\
4   &   Citeomatic \shortcite{citeomatic}              & 10.11          & 15.91          & 26.34          & 18.59          & 8.82          & 15.07          & 22.88          & 14.96           \\

5   &   H-Transformer \shortcite{lcr}                  & 8.42              & 15.11              & 22.67          & 21.37          & 4.40              & 10.85              & 10.12          & 9.59            \\
6   &   BERT \shortcite{bert}                          & 11.81          & 19.60          & 29.86          & 20.20          & 8.48          & 14.38          & 25.60          & 15.05           \\
7   &   SciBERT \shortcite{scibert}                    & 13.27          & 21.17          & 31.69          & \underline{23.55}          & 9.72          & 16.44          & 27.47          & \underline{17.94}           \\
8   &   SPECTER \shortcite{specter}                    & \underline{13.44}          & \underline{21.87}          & \underline{31.92}          & 23.35          & \underline{9.83}          & \underline{16.60}          & \underline{27.48}          & 17.47           \\
9   &   CiteBERT \shortcite{citebert}                  & 12.70          & 20.61          & 31.23          & 23.15          & 9.19          & 15.44          & 26.91          & 17.03           \\

10   &   LinkBERT \shortcite{linkbert}                  & 12.75          & 21.15          & 31.39          & 22.86          & 8.86          & 15.32          & 26.52          & 17.03           \\
11   &   SciNCL \shortcite{scincl}                      & 13.19          & 21.01          & 31.64          & 22.38          & 9.83          & 16.60          & 27.48          & 17.47           \\

12   &   \framename \ (ours)                               & \textbf{14.94} & \textbf{23.13} & \textbf{33.28} & \textbf{25.59} & \textbf{10.48} & \textbf{17.14} & \textbf{28.28} & \textbf{18.76}  \\ 
\hdashline
  &  $\pm \sigma$ w/ five seeds                        & \textit{.185}           & \textit{.330}           & \textit{.168}          & \textit{.541}           & \textit{.241}           & \textit{.422}           & \textit{.219}          & \textit{.369}            \\
\bottomrule[1pt]
\end{tabular}
% }
\caption{Performance results on \dsname \ and Extended \dsname. Our scores are reported as mean and standard deviation $\sigma$ over five random seeds.}
  \label{tab:main}%
\end{table*}

\subsection{Approaches for Comparison}\label{sec:baselines}
We have four groups of approaches for comparison.

The first group includes: 

(1-1) \textbf{BM25} \cite{bm25}, which is a highly effective strong baseline model representing traditional sparse retrieval models.

The second group comprises two citation recommendation models:

(2-1) \textbf{Citeomatic} \cite{citeomatic}, a global CR model uses shallow feed-forward networks to learn representations of papers.

(2-2) \textbf{H-Transformer} \cite{lcr}, a local CR model adopts hierarchical transformer layers as paper encoders in the prefetching stage and generates recommendations by pair-wise reranking via SciBERT.

The third group consists of six pre-trained scientific document encoding models:

(3-1) \textbf{BERT} \cite{bert}, the dominant pre-trained model which achieves great success on various language understanding tasks.

(3-2) \textbf{SciBERT} \cite{scibert}, a variant of BERT trained on a corpus of scientific articles with masked language modeling objectives.

(3-3) \textbf{SPECTER} \cite{specter}, a SciBERT-based scientific document encoder trained with a contrastive learning objective that minimizes the L2 distance between embeddings of citing-cited paper pairs.

(3-4) \textbf{CiteBERT} \cite{citebert}, a variant of SciBERT fine-tuned on cite-worthiness detection task.

(3-5) \textbf{LinkBERT} \cite{linkbert}, a variant of BERT fine-tuned on document relation prediction task.

(3-6) \textbf{SciNCL} \cite{scincl}, a SciBERT-based Encoder that uses citation graph embeddings for a more informative selection of negative examples.

The fourth group includes two large language models:

(4-1) \textbf{ChatGPT-turbo}, a representative, powerful and widely used conversational agent.

(4-2) \textbf{LLaMA2-13b} \cite{llama2}, a popular advanced large language model.

The input of baselines (1-1)-(3-6) is the concatenation of the title, abstract, and titles in the reference section of the input paper.
For LLMs, we design prompts similar to those used in news recommendation \cite{Liu2023ONCEBC} and let LLMs rerank the top ten\footnote{We also tested 20 and 30, but larger numbers resulted in slightly worse performance.} results of BM25.
While baselines (2-1), (2-2), (4-1), and (4-2) directly output a sorted candidates list, other baselines generate vectors of papers upon which we calculate similarities and rank candidates.

\subsection{Metrics and Implementation Details}
Following previous work, we use four commonly used evaluation metrics: 
(1) Mean Average Precision (\textbf{MAP});
(2) Mean Reciprocal Rank (\textbf{MRR}) \cite{mrr};
(3) Normalized Discounted Cumulative Gain (\textbf{NDCG}) \cite{ndcg}, a widely used measure of ranking quality and is computed by  
\begin{equation}
    {\rm NDCG} = \sum\limits_{i=1}^{|M|} \frac{2^{r(i)}-1}{log_2(i+1)}
\end{equation}{}
where $M$ is the sorted list of papers output by models and $r(i) = 1$ if the $i$-th paper is the one reviewers recommended to cite, otherwise $r(i) = 0$.  
(4) \textbf{Recall@K}.
Considering the scale of \dsname, we set $K=10$.

We adopt the weights of SciNCL \cite{scincl} to initialize our text encoder and the loss margin $m$ is set as 0.05 following \cite{lcr}.
The optimizer is AdamW \cite{Loshchilov2017FixingWD} with a learning rate of $\lambda = 2^{-5}$.
Models are trained on a single NVIDIA GeForce V100 (32GB) GPU for five epochs, and its checkpoint is updated every $N_{iter}=5,000$. 
We set $\alpha = 0.6$, $K_{n}=100$ and the ratio of positive/hard negative/easy negative papers to 1:1:1.

\subsection{Main Results}
The main results are summarized in Table \ref{tab:main}.
The overall best and previously best results are boldfaced and underlined, respectively.
% As summarized in Table \ref{tab:main}, the overall best and previously best results are boldfaced and underlined, respectively.
We have several observations:

Firstly, our proposed model achieves the best results in terms of all metrics on two versions of \dsname, displaying its superiority to other methods.

Secondly, pre-trained methods (Rows 6-12) perform generally better than other methods. 
The deeply stacked and large-scale pre-trained BERT model can better model text semantics than shallow word embeddings, which is crucial for content understanding in recommending citations.
For H-Transformer, although it adopts hierarchical transformer layers as paper encoders, the lack of pre-training on large-scale corpora makes it fail in \taskname. In contrast, Citeomatic obtains better results with shallow networks sufficiently trained on the dataset.

% Thirdly, compared with BERT and SciBERT, LMs with citation informed (Rows 8-12) during training generally achieve better results. This is straightforward since the task requires two documents connected with citations to be similar in the embedding space.

% Thirdly, our proposed model outperforms all other baseline methods in terms of all metrics on the two versions of \dsname. The improvements may be attributed to the transformer-based LM pre-trained on a large corpus, the \textit{Attentive Reference Encoder} part mining relations between input papers and their already cited papers, and contrastive learning objectives for effectively training the model.

Thirdly, LLMs exhibit mediocre performance on \taskname, with a notable disparity when compared to other BERT-based models. 
The results of reranking with LLaMA2-13b have even become worse on Extended \dsname.
It seems that reranking is a complex task that goes beyond the capabilities of LLMs alone, which are good at generating texts and understanding contexts.

Fourthly, while H-Transformer achieves \textbf{75.7\%} at R@10 on local CR \cite{lcr} and Citeomatic obtains \textbf{77.1\%} at MRR on global CR \cite{citeomatic}, neither model achieves more than 30\% at any metrics on \taskname.
Apparently, existing models have much worse performance on \taskname \ compared to traditional CR datasets, highlighting the complexity of \taskname, and the need for more sophisticated recommendation methods.

\begin{figure*}[ht] 
\centering
    \begin{minipage}[t]{0.48\textwidth}
    \centering
    \includegraphics[width=0.95\linewidth]{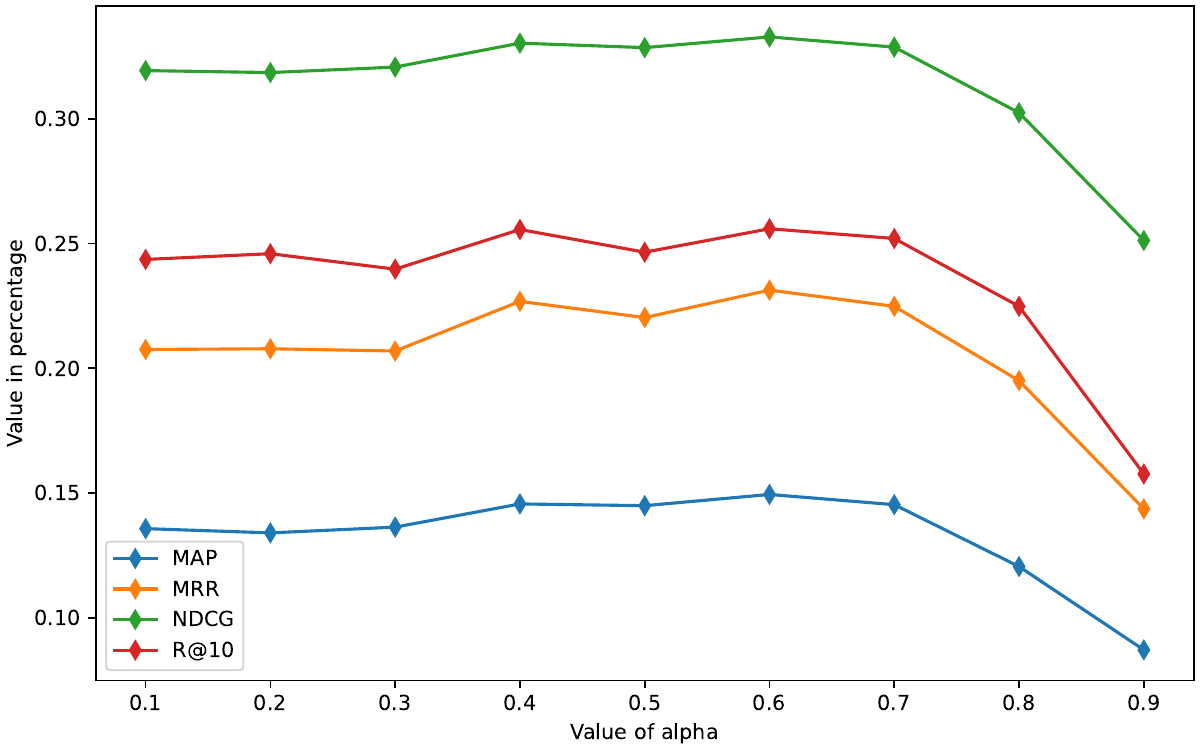}
    \caption{Results with different values of $\alpha$. }
    \label{fig:aplha}
    \end{minipage} 
    \hfill
    \begin{minipage}[t]{0.48\textwidth}
    \centering
    \includegraphics[width=0.95\linewidth]{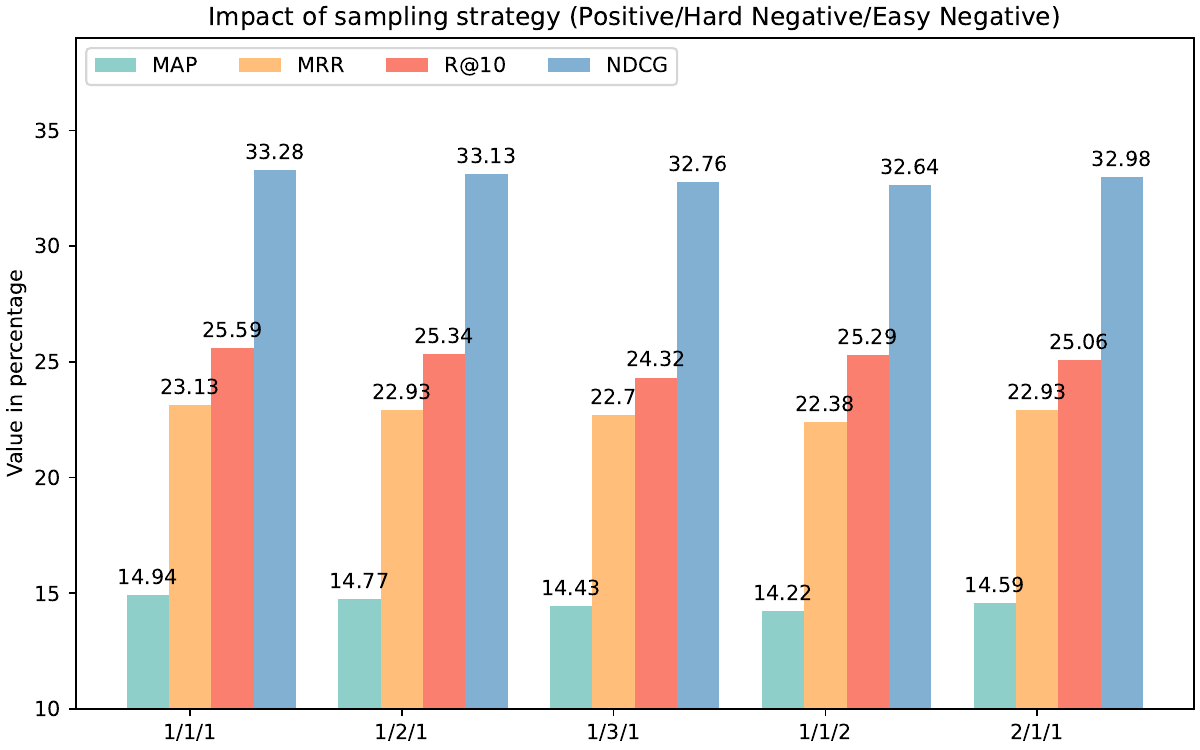}
    \caption{Results with different sampling strategies. }
    \label{fig:sampling}
    \end{minipage}
\end{figure*}

\subsection{Ablation Study}

In this section, we study the effectiveness of different components of our proposed model by removing or replacing them.
% The results on \dsname \ are illustrated in Table \ref{tab:ablation}.

We first show the effect of the \textit{Attentive Reference Encoder} part by removing it. 
% Regardless of the choice of text encoders, \textit{Attentive Reference Encoder} is the key structural difference between our framework and CR models. 
The results in Table \ref{tab:ablation} demonstrate that \textit{ARE} plays a crucial role, indicating that it is necessary to mine the citing patterns by modeling already cited papers in \taskname.
We also remove the hard negative examples when training the model.
The decreased performance verifies the benefits of training the model to distinguish hard negative examples.
Nevertheless, the results of the model with both components removed are between those of the model with only one component removed. 
It implies that \textit{ARE} has a greater impact than hard negative examples.

To further examine the effect of the \textit{ARE}, we explore some other ways to tackle the reference section.
``average pooling'' means replacing the \textit{Attention} layer with an average layer when calculating the representation of the reference ${\boldsymbol {\rm v}}_{R}$.
``concatenation'' means replacing the linear combination with concatenation when calculating the final representation of the input paper ${\boldsymbol {\rm v}}_{p}$.
However, these replaced models result in a decrease in performance, proving the superiority of balancing the content and citation pattern in our method.

% Table generated by Excel2LaTeX from sheet 'Sheet4'
\begin{table}[t]
\centering
\resizebox{0.98\linewidth}{!}{
\begin{tabular}{l|cccc}
    \toprule
    Model & MAP   & MRR  & NDCG & R@10 \\
    \midrule
    Ours  & \textbf{14.94} & \textbf{23.13} & \textbf{33.28} & \textbf{25.59} \\
     \quad w/o ARE & 13.10 & 20.17 & 31.39 & 23.03 \\
     \quad w/o Hard Negatives (HN) & 14.36 & 22.45 & 32.70 & 25.13 \\
     \quad w/o ARE \& w/o HN & 13.61 & 20.86 & 31.90 & 24.45 \\
    \hdashline
    %  \quad words concatenation & 13.26 & 21.30 & 31.72 & 22.57 \\
     \quad average pooling & 12.57 & 19.75 & 30.73 & 21.52 \\
     \quad concatenation & 13.68 & 21.25 & 32.11 & 24.35 \\
    \bottomrule
    \end{tabular}%
}
 \caption{Ablation results on \dsname}
  \label{tab:ablation}%
\end{table}%

\subsection{Parameter Analysis}

\subsubsection{Effect of $\alpha$}
Figure \ref{fig:aplha} displays the results on \dsname \ of different values of $\alpha$, which is the linear weight of combining ${\textbf{v}_{\rm content}}$ and ${\textbf{v}_{\rm R}}$.
The performance of the model achieves the best when $\alpha = 0.6$ and sharply decreases when $\alpha$ exceeds $0.7$.

\subsubsection{Effect of Sampling Strategy}
Figure \ref{fig:sampling} presents the results on \dsname \ of different sampling strategies. 
Here a strategy means the number of positive/hard negative/easy negative papers in a training instance.
Considering the statistics of \dsname, we set the number of papers in an training instance at a low level.
For all metrics, the best results are achieved by strategy ``1/1/1". 
Increasing the number of any kind of papers in a training instance raises a slight decrease in performance. 
This may be due to the current scale of \dsname.

\section{Challenges and Future Work}

In this section, we introduce the challenges and possible further research directions of \taskname \ from three aspects:

% \textbf{Relevance:}
\subsection{Relevance}
One key aspect of \taskname \ is measuring the relevance between missed citations and submissions. 
We have designed an \textit{ARE} module that leverages the capability of \textit{Attention} to exploit potential relevance. 
However, along with most existing methods, our method solely considers the title and abstract of a scientific paper but ignores its body text.
One possible research direction is incorporating additional information from papers in modeling while ensuring efficiency. 
Relevance may be more comprehensively explored with more information provided, especially the body text of submissions \citep{jcdl13, yenkan15}.
However, accurately and effectively modeling such long texts and mining the relevance that may only be mentioned in a short context is still challenging.

% \textbf{Intent:}
\subsection{Intent}
 Both previous methods and our method focus on mining relevance, assuming that relevance is implicit in the text. 
 However, cases may be that the citations recommended by reviewers are completely ignored by the submissions, and there is no textual relevance. 
 In such cases, we believe that mining the intent behind the reviewers' recommendations can be helpful. 
 As shown in Figure \ref{fig:reviews}, reviewers often provide explanations for their recommended citations. 
 Therefore, a feasible research direction is to automatically locate these explanations from review texts and utilize them to train smarter recommenders.
 
% \textbf{Understanding:}
\subsection{Understanding}
Generally, reviewers rely on their accumulated academic knowledge and understanding of submissions to identify missed citations. 
Meanwhile, LLMs have shown remarkable capabilities in language understanding \cite{Yang2023HarnessingTP} and can effectively apply their learned knowledge and reasoning abilities to tackle new tasks \cite{Zhu2023LargeLM}.
Thus, one promising direction is to leverage the power of LLMs to obtain a fine-grained understanding of submissions and identify potential weaknesses that need further citations to support.

\section{Conclusion}
In this paper, we introduce a novel challenging task of \underline{R}ecommending \underline{M}issed \underline{C}itations Identified by Reviewers (\taskname).
\taskname \ aims to improve the citations of submissions and avoid the reliability and novelty of research being undermined because of missing vital citations.
We curate a high-quality dataset named \dsname \ by extracting submission-citation pairs labeled by reviewers from real reviews in top-tier conferences.
We conduct an extensive evaluation of four groups of strong methods on the developed dataset.
Moreover, we propose a novel framework \framename \ by integrating an \textit{Attentive Reference Encoder} module and contrastive learning objectives.
Our proposed method outperforms all baselines in all metrics and can serve as a strong baseline.
We highlight challenges and several potential research directions of \taskname.
We make all code and data publicly available to motivate future research.

\section*{Ethics Statement}
The datasets used in our research are collected through open-source approaches.
The whole process is conducted legally, following ethical requirements.

We see opportunities for researchers to apply our built datasets and models to help identify missed citations of academic manuscripts. 
However, machine learning models are liable to amplify biases in training data \cite{Hall2022ASS}, and current collected datasets are limited in scale and diversity.
Researchers must consider these implications when conducting work on our datasets and consider whether a recommended citation is feasible when using models in practice.

\clearpage

\section*{References}\label{sec:reference}
\normalem
\bibliographystyle{lrec_natbib}
\bibliography{reference}

\begin{thebibliography}{66}
\expandafter\ifx\csname natexlab\endcsname\relax\def\natexlab#1{#1}\fi

\bibitem[{Arous et~al.(2021)Arous, Yang, Khayati, and
  Cudr{\'e}-Mauroux}]{re_arous_21}
Ines Arous, Jie Yang, Mourad Khayati, and Philippe Cudr{\'e}-Mauroux. 2021.
\newblock Peer grading the peer reviews: A dual-role approach for lightening
  the scholarly paper review process.
\newblock \emph{Proceedings of the Web Conference 2021}.

\bibitem[{Beltagy et~al.(2019)Beltagy, Lo, and Cohan}]{scibert}
Iz~Beltagy, Kyle Lo, and Arman Cohan. 2019.
\newblock Scibert: A pretrained language model for scientific text.
\newblock In \emph{Conference on Empirical Methods in Natural Language
  Processing}.

\bibitem[{Bhagavatula et~al.(2018)Bhagavatula, Feldman, Power, and
  Ammar}]{citeomatic}
Chandra Bhagavatula, Sergey Feldman, Russell Power, and Waleed Ammar. 2018.
\newblock Content-based citation recommendation.
\newblock In \emph{North American Chapter of the Association for Computational
  Linguistics}.

\bibitem[{Bharti et~al.(2022)Bharti, Ghosal, Agarwal, and
  Ekbal}]{Bharti2022AMF}
Prabhat~Kumar Bharti, Tirthankar Ghosal, Mayank Agarwal, and Asif Ekbal. 2022.
\newblock A method for automatically estimating the informativeness of peer
  reviews.
\newblock In \emph{ICON}.

\bibitem[{Booth et~al.(2012)Booth, Papaioannou, and Sutton}]{booklr}
Andrew Booth, Diana Papaioannou, and Anthea Sutton. 2012.
\newblock \emph{Systematic Approaches to a Successful Literature Review}.

\bibitem[{Brown et~al.(2020)Brown, Mann, Ryder, Subbiah, Kaplan, Dhariwal,
  Neelakantan, Shyam, Sastry, Askell, Agarwal, Herbert{-}Voss, Krueger,
  Henighan, Child, Ramesh, Ziegler, Wu, Winter, Hesse, Chen, Sigler, Litwin,
  Gray, Chess, Clark, Berner, McCandlish, Radford, Sutskever, and
  Amodei}]{gpt3}
Tom~B. Brown, Benjamin Mann, Nick Ryder, Melanie Subbiah, Jared Kaplan,
  Prafulla Dhariwal, Arvind Neelakantan, Pranav Shyam, Girish Sastry, Amanda
  Askell, Sandhini Agarwal, Ariel Herbert{-}Voss, Gretchen Krueger, Tom
  Henighan, Rewon Child, Aditya Ramesh, Daniel~M. Ziegler, Jeffrey Wu, Clemens
  Winter, Christopher Hesse, Mark Chen, Eric Sigler, Mateusz Litwin, Scott
  Gray, Benjamin Chess, Jack Clark, Christopher Berner, Sam McCandlish, Alec
  Radford, Ilya Sutskever, and Dario Amodei. 2020.
\newblock Language models are few-shot learners.
\newblock In \emph{Advances in Neural Information Processing Systems 33: Annual
  Conference on Neural Information Processing Systems 2020, NeurIPS 2020,
  December 6-12, 2020, virtual}.

\bibitem[{Cai et~al.(2018)Cai, Han, and Yang}]{cai_generative_2018}
Xiaoyan Cai, Junwei Han, and Libin Yang. 2018.
\newblock Generative adversarial network based heterogeneous bibliographic
  network representation for personalized citation recommendation.
\newblock In \emph{AAAI}.

\bibitem[{Cai et~al.(2022)Cai, Liu, Yang, Lu, Zhao, Shen, and
  Liu}]{Cai2022COVIDSumAL}
Xiaoyan Cai, Sen Liu, Libin Yang, Yan Lu, Jintao Zhao, Dinggang Shen, and
  Tianming Liu. 2022.
\newblock Covidsum: A linguistically enriched scibert-based summarization model
  for covid-19 scientific papers.
\newblock \emph{Journal of Biomedical Informatics}, 127:103999 -- 103999.

\bibitem[{Chakraborty et~al.(2020)Chakraborty, Goyal, and
  Mukherjee}]{re_chak_20}
Souvic Chakraborty, Pawan Goyal, and Animesh Mukherjee. 2020.
\newblock Aspect-based sentiment analysis of scientific reviews.
\newblock \emph{Proceedings of the ACM/IEEE Joint Conference on Digital
  Libraries in 2020}.

\bibitem[{Chen et~al.(2021)Chen, Tworek, Jun, Yuan, Ponde, Kaplan, Edwards,
  Burda, Joseph, Brockman, Ray, Puri, Krueger, Petrov, Khlaaf, Sastry, Mishkin,
  Chan, Gray, Ryder, Pavlov, Power, Kaiser, Bavarian, Winter, Tillet, Such,
  Cummings, Plappert, Chantzis, Barnes, Herbert-Voss, Guss, Nichol, Babuschkin,
  Balaji, Jain, Carr, Leike, Achiam, Misra, Morikawa, Radford, Knight,
  Brundage, Murati, Mayer, Welinder, McGrew, Amodei, McCandlish, Sutskever, and
  Zaremba}]{codex}
Mark Chen, Jerry Tworek, Heewoo Jun, Qiming Yuan, Henrique Ponde, Jared Kaplan,
  Harrison Edwards, Yura Burda, Nicholas Joseph, Greg Brockman, Alex Ray, Raul
  Puri, Gretchen Krueger, Michael Petrov, Heidy Khlaaf, Girish Sastry, Pamela
  Mishkin, Brooke Chan, Scott Gray, Nick Ryder, Mikhail Pavlov, Alethea Power,
  Lukasz Kaiser, Mohammad Bavarian, Clemens Winter, Philippe Tillet,
  Felipe~Petroski Such, David~W. Cummings, Matthias Plappert, Fotios Chantzis,
  Elizabeth Barnes, Ariel Herbert-Voss, William~H. Guss, Alex Nichol, Igor
  Babuschkin, S.~Arun Balaji, Shantanu Jain, Andrew Carr, Jan Leike, Joshua
  Achiam, Vedant Misra, Evan Morikawa, Alec Radford, Matthew~M. Knight, Miles
  Brundage, Mira Murati, Katie Mayer, Peter Welinder, Bob McGrew, Dario Amodei,
  Sam McCandlish, Ilya Sutskever, and Wojciech Zaremba. 2021.
\newblock Evaluating large language models trained on code.
\newblock \emph{ArXiv}, abs/2107.03374.

\bibitem[{Cohan et~al.(2020)Cohan, Feldman, Beltagy, Downey, and
  Weld}]{specter}
Arman Cohan, Sergey Feldman, Iz~Beltagy, Doug Downey, and Daniel~S. Weld. 2020.
\newblock Specter: Document-level representation learning using
  citation-informed transformers.
\newblock In \emph{Annual Meeting of the Association for Computational
  Linguistics}.

\bibitem[{Deng et~al.(2020)Deng, Peng, Xia, Li, He, and Yu}]{re_deng_20}
Zhongfen Deng, Hao Peng, Congying Xia, Jianxin Li, Lifang He, and Philip~S. Yu.
  2020.
\newblock Hierarchical bi-directional self-attention networks for paper review
  rating recommendation.
\newblock \emph{ArXiv}, abs/2011.00802.

\bibitem[{Devlin et~al.(2019)Devlin, Chang, Lee, and Toutanova}]{bert}
Jacob Devlin, Ming-Wei Chang, Kenton Lee, and Kristina Toutanova. 2019.
\newblock Bert: Pre-training of deep bidirectional transformers for language
  understanding.
\newblock \emph{ArXiv}, abs/1810.04805.

\bibitem[{Ebesu and Fang(2017)}]{Ebesu2017NeuralCN}
Travis Ebesu and Yi~Fang. 2017.
\newblock Neural citation network for context-aware citation recommendation.
\newblock \emph{Proceedings of the 40th International ACM SIGIR Conference on
  Research and Development in Information Retrieval}.

\bibitem[{Fan et~al.(2023)Fan, Zhao, Li, Liu, Mei, Wang, Tang, and
  Li}]{Fan2023RecommenderSI}
Wenqi Fan, Zihuai Zhao, Jiatong Li, Yunqing Liu, Xiaowei Mei, Yiqi Wang,
  Jiliang Tang, and Qing Li. 2023.
\newblock Recommender systems in the era of large language models (llms).
\newblock \emph{ArXiv}, abs/2307.02046.

\bibitem[{F{\"a}rber and Jatowt(2020)}]{review2020}
Michael F{\"a}rber and Adam Jatowt. 2020.
\newblock Citation recommendation: approaches and datasets.
\newblock \emph{International Journal on Digital Libraries}, pages 1 -- 31.

\bibitem[{F{\"a}rber and Sampath(2020)}]{hybrid20}
Michael F{\"a}rber and Ashwath Sampath. 2020.
\newblock Hybridcite: A hybrid model for context-aware citation recommendation.
\newblock \emph{Proceedings of the ACM/IEEE Joint Conference on Digital
  Libraries in 2020}.

\bibitem[{Gu et~al.(2021)Gu, Gao, and Hahnloser}]{lcr}
Nianlong Gu, Yingqiang Gao, and Richard H.~R. Hahnloser. 2021.
\newblock Local citation recommendation with hierarchical-attention text
  encoder and scibert-based reranking.
\newblock In \emph{ECIR}.

\bibitem[{Gupta and Varma(2017)}]{gupta_scientific_2017}
Shashank Gupta and Vasudeva Varma. 2017.
\newblock Scientific article recommendation by using distributed
  representations of text and graph.
\newblock \emph{Proceedings of the 26th International Conference on World Wide
  Web Companion}.

\bibitem[{Hall et~al.(2022)Hall, van~der Maaten, Gustafson, and
  Adcock}]{Hall2022ASS}
Melissa Hall, Laurens van~der Maaten, Laura Gustafson, and Aaron~B. Adcock.
  2022.
\newblock A systematic study of bias amplification.
\newblock \emph{ArXiv}, abs/2201.11706.

\bibitem[{Han et~al.(2018)Han, Song, Zhao, Shi, and Zhang}]{acl18}
Jialong Han, Yan Song, Wayne~Xin Zhao, Shuming Shi, and Haisong Zhang. 2018.
\newblock hyperdoc2vec: Distributed representations of hypertext documents.
\newblock In \emph{ACL}.

\bibitem[{He et~al.(2011)He, Kifer, Pei, Mitra, and Giles}]{He2011CitationRW}
Qi~He, Daniel Kifer, Jian Pei, Prasenjit Mitra, and C.~Lee Giles. 2011.
\newblock Citation recommendation without author supervision.
\newblock In \emph{WSDM '11}.

\bibitem[{He et~al.(2010)He, Pei, Kifer, Mitra, and
  Giles}]{He2010ContextawareCR}
Qi~He, Jian Pei, Daniel Kifer, Prasenjit Mitra, and C.~Lee Giles. 2010.
\newblock Context-aware citation recommendation.
\newblock In \emph{WWW '10}.

\bibitem[{Hou et~al.(2023)Hou, Zhang, Lin, Lu, Xie, McAuley, and
  Zhao}]{Hou2023LargeLM}
Yupeng Hou, Junjie Zhang, Zihan Lin, Hongyu Lu, Ruobing Xie, Julian McAuley,
  and Wayne~Xin Zhao. 2023.
\newblock Large language models are zero-shot rankers for recommender systems.
\newblock \emph{ArXiv}, abs/2305.08845.

\bibitem[{Hu et~al.(2020)Hu, He, Tan, Zhang, and Ge}]{Hu2020FusionOD}
Yanli Hu, Chunhui He, Zhen Tan, Chong Zhang, and Bin Ge. 2020.
\newblock Fusion of domain knowledge and text features for query expansion in
  citation recommendation.
\newblock In \emph{KSEM}.

\bibitem[{Hua et~al.(2019)Hua, Nikolov, Badugu, and Wang}]{re_hua_19}
Xinyu Hua, Mitko Nikolov, Nikhil Badugu, and Lu~Wang. 2019.
\newblock Argument mining for understanding peer reviews.
\newblock \emph{ArXiv}, abs/1903.10104.

\bibitem[{Huang et~al.(2012)Huang, Kataria, Caragea, Mitra, Giles, and
  Rokach}]{Huang2012RecommendingCT}
Wenyi Huang, Saurabh Kataria, Cornelia Caragea, Prasenjit Mitra, C.~Lee Giles,
  and Lior Rokach. 2012.
\newblock Recommending citations: translating papers into references.
\newblock \emph{Proceedings of the 21st ACM international conference on
  Information and knowledge management}.

\bibitem[{Huang et~al.(2015)Huang, Wu, Liang, Mitra, and Giles}]{Huang2015ANP}
Wenyi Huang, Zhaohui Wu, Chen Liang, Prasenjit Mitra, and C.~Lee Giles. 2015.
\newblock A neural probabilistic model for context based citation
  recommendation.
\newblock In \emph{AAAI}.

\bibitem[{J{\"{a}}rvelin and Kek{\"{a}}l{\"{a}}inen(2002)}]{ndcg}
Kalervo J{\"{a}}rvelin and Jaana Kek{\"{a}}l{\"{a}}inen. 2002.
\newblock Cumulated gain-based evaluation of {IR} techniques.
\newblock \emph{{ACM} Trans. Inf. Syst.}, 20(4):422--446.

\bibitem[{Jiang et~al.(2018)Jiang, Yin, Gao, Lu, and
  Liu}]{jiang_cross-language_2018}
Zhuoren Jiang, Yue Yin, Liangcai Gao, Yao Lu, and Xiaozhong Liu. 2018.
\newblock Cross-language citation recommendation via hierarchical
  representation learning on heterogeneous graph.
\newblock \emph{The 41st International ACM SIGIR Conference on Research \&
  Development in Information Retrieval}.

\bibitem[{Kumar et~al.(2022)Kumar, Arora, Ghosal, and Ekbal}]{re_kumar_22}
Sandeep Kumar, Hardik Arora, Tirthankar Ghosal, and Asif Ekbal. 2022.
\newblock Deepaspeer: Towards an aspect-level sentiment controllable framework
  for decision prediction from academic peer reviews.
\newblock \emph{2022 ACM/IEEE Joint Conference on Digital Libraries (JCDL)},
  pages 1--11.

\bibitem[{Li et~al.(2020)Li, Sato, Shimura, and Fukumoto}]{re_li_20}
Jiyi Li, Ayaka Sato, Kazuya Shimura, and Fumiyo Fukumoto. 2020.
\newblock Multi-task peer-review score prediction.
\newblock In \emph{SDP}.

\bibitem[{Li et~al.(2022)Li, Li, Zhao, Ding, and rong Wen}]{re_li_22}
Siqing Li, Yaliang Li, Wayne~Xin Zhao, Bolin Ding, and Ji~rong Wen. 2022.
\newblock Interpretable aspect-aware capsule network for peer review based
  citation count prediction.
\newblock \emph{ACM Transactions on Information Systems}.

\bibitem[{Li et~al.(2019)Li, Zhao, Yin, and rong Wen}]{re_li_19}
Siqing Li, Wayne~Xin Zhao, Eddy~Jing Yin, and Ji~rong Wen. 2019.
\newblock A neural citation count prediction model based on peer review text.
\newblock In \emph{Conference on Empirical Methods in Natural Language
  Processing}.

\bibitem[{Lin et~al.(2021)Lin, Song, Zhou, Chen, and Shi}]{re_lin_21}
Jialiang Lin, Jiaxin Song, Zhangping Zhou, Yidong Chen, and Xiaodon Shi. 2021.
\newblock Automated scholarly paper review: Technologies and challenges.

\bibitem[{Liu et~al.(2023)Liu, Chen, Sakai, and Wu}]{Liu2023ONCEBC}
Qijiong Liu, Nuo Chen, Tetsuya Sakai, and Xiao-Ming Wu. 2023.
\newblock Once: Boosting content-based recommendation with both open- and
  closed-source large language models.

\bibitem[{Liu et~al.(2014)Liu, Yu, Guo, and Sun}]{cikm14}
Xiaozhong Liu, Yingying Yu, Chun Guo, and Yizhou Sun. 2014.
\newblock Meta-path-based ranking with pseudo relevance feedback on
  heterogeneous graph for citation recommendation.
\newblock In \emph{Proceedings of the 23rd {ACM} International Conference on
  Conference on Information and Knowledge Management, {CIKM} 2014, Shanghai,
  China, November 3-7, 2014}, pages 121--130. {ACM}.

\bibitem[{Loshchilov and Hutter(2017)}]{Loshchilov2017FixingWD}
Ilya Loshchilov and Frank Hutter. 2017.
\newblock Fixing weight decay regularization in adam.
\newblock \emph{ArXiv}, abs/1711.05101.

\bibitem[{Luu et~al.(2020)Luu, Wu, Koncel-Kedziorski, Lo, Cachola, and
  Smith}]{Luu2020ExplainingRB}
Kelvin Luu, Xinyi Wu, Rik Koncel-Kedziorski, Kyle Lo, Isabel Cachola, and
  Noah~A. Smith. 2020.
\newblock Explaining relationships between scientific documents.
\newblock In \emph{Annual Meeting of the Association for Computational
  Linguistics}.

\bibitem[{Medic and Snajder(2022)}]{Medic2022LargescaleEO}
Zoran Medic and Jan Snajder. 2022.
\newblock Large-scale evaluation of transformer-based article encoders on the
  task of citation recommendation.
\newblock In \emph{SDP}.

\bibitem[{Ostendorff et~al.(2022)Ostendorff, Rethmeier, Augenstein, Gipp, and
  Rehm}]{scincl}
Malte Ostendorff, Nils Rethmeier, Isabelle Augenstein, Bela Gipp, and Georg
  Rehm. 2022.
\newblock Neighborhood contrastive learning for scientific document
  representations with citation embeddings.
\newblock \emph{ArXiv}, abs/2202.06671.

\bibitem[{Pears and Shields(2013)}]{bookcite}
Richard Pears and Graham Shields. 2013.
\newblock \emph{Cite Them Right: The Essential Referencing Guide}.

\bibitem[{Pornprasit et~al.(2022)Pornprasit, Liu, Kiattipadungkul,
  Kertkeidkachorn, Kim, Noraset, Hassan, and
  Tuarob}]{Pornprasit2022EnhancingCR}
Chanathip Pornprasit, Xin Liu, Pattararat Kiattipadungkul, Natthawut
  Kertkeidkachorn, Kyoung-Sook Kim, Thanapon Noraset, Saeed-Ul Hassan, and
  Suppawong Tuarob. 2022.
\newblock Enhancing citation recommendation using citation network embedding.
\newblock \emph{Scientometrics}, 127:233 -- 264.

\bibitem[{Ren et~al.(2014)Ren, Liu, Yu, Khandelwal, Gu, Wang, and
  Han}]{Ren2014ClusCiteEC}
Xiang Ren, Jialu Liu, Xiao Yu, Urvashi Khandelwal, Quanquan Gu, Lidan Wang, and
  Jiawei Han. 2014.
\newblock Cluscite: effective citation recommendation by information
  network-based clustering.
\newblock \emph{Proceedings of the 20th ACM SIGKDD international conference on
  Knowledge discovery and data mining}.

\bibitem[{Robertson and Walker(1999)}]{bm25}
Stephen~E. Robertson and Steve Walker. 1999.
\newblock Okapi/keenbow at trec-8.
\newblock In \emph{Text Retrieval Conference}.

\bibitem[{Sugiyama and Kan(2013)}]{jcdl13}
Kazunari Sugiyama and Min{-}Yen Kan. 2013.
\newblock Exploiting potential citation papers in scholarly paper
  recommendation.
\newblock In \emph{13th {ACM/IEEE-CS} Joint Conference on Digital Libraries,
  {JCDL} '13, Indianapolis, IN, USA, July 22 - 26, 2013}, pages 153--162.
  {ACM}.

\bibitem[{Sugiyama and Kan(2015)}]{yenkan15}
Kazunari Sugiyama and Min{-}Yen Kan. 2015.
\newblock A comprehensive evaluation of scholarly paper recommendation using
  potential citation papers.
\newblock \emph{Int. J. Digit. Libr.}, 16(2):91--109.

\bibitem[{Tang and Zhang(2009)}]{Tang2009ADA}
Jie Tang and Jing Zhang. 2009.
\newblock A discriminative approach to topic-based citation recommendation.
\newblock In \emph{PAKDD}.

\bibitem[{Touvron et~al.(2023{\natexlab{a}})Touvron, Lavril, Izacard, Martinet,
  Lachaux, Lacroix, Rozi{\`e}re, Goyal, Hambro, Azhar, Rodriguez, Joulin,
  Grave, and Lample}]{llama}
Hugo Touvron, Thibaut Lavril, Gautier Izacard, Xavier Martinet, Marie-Anne
  Lachaux, Timoth{\'e}e Lacroix, Baptiste Rozi{\`e}re, Naman Goyal, Eric
  Hambro, Faisal Azhar, Aurelien Rodriguez, Armand Joulin, Edouard Grave, and
  Guillaume Lample. 2023{\natexlab{a}}.
\newblock Llama: Open and efficient foundation language models.
\newblock \emph{ArXiv}, abs/2302.13971.

\bibitem[{Touvron et~al.(2023{\natexlab{b}})Touvron, Martin, Stone, Albert,
  Almahairi, Babaei, Bashlykov, Batra, Bhargava, Bhosale, Bikel, Blecher,
  Ferrer, Chen, Cucurull, Esiobu, Fernandes, Fu, Fu, Fuller, Gao, Goswami,
  Goyal, Hartshorn, Hosseini, Hou, Inan, Kardas, Kerkez, Khabsa, Kloumann,
  Korenev, Koura, Lachaux, Lavril, Lee, Liskovich, Lu, Mao, Martinet, Mihaylov,
  Mishra, Molybog, Nie, Poulton, Reizenstein, Rungta, Saladi, Schelten, Silva,
  Smith, Subramanian, Tan, Tang, Taylor, Williams, Kuan, Xu, Yan, Zarov, Zhang,
  Fan, Kambadur, Narang, Rodriguez, Stojnic, Edunov, and Scialom}]{llama2}
Hugo Touvron, Louis Martin, Kevin~R. Stone, Peter Albert, Amjad Almahairi,
  Yasmine Babaei, Nikolay Bashlykov, Soumya Batra, Prajjwal Bhargava, Shruti
  Bhosale, Daniel~M. Bikel, Lukas Blecher, Cristian~Cant{\'o}n Ferrer, Moya
  Chen, Guillem Cucurull, David Esiobu, Jude Fernandes, Jeremy Fu, Wenyin Fu,
  Brian Fuller, Cynthia Gao, Vedanuj Goswami, Naman Goyal, Anthony~S.
  Hartshorn, Saghar Hosseini, Rui Hou, Hakan Inan, Marcin Kardas, Viktor
  Kerkez, Madian Khabsa, Isabel~M. Kloumann, A.~V. Korenev, Punit~Singh Koura,
  Marie-Anne Lachaux, Thibaut Lavril, Jenya Lee, Diana Liskovich, Yinghai Lu,
  Yuning Mao, Xavier Martinet, Todor Mihaylov, Pushkar Mishra, Igor Molybog,
  Yixin Nie, Andrew Poulton, Jeremy Reizenstein, Rashi Rungta, Kalyan Saladi,
  Alan Schelten, Ruan Silva, Eric~Michael Smith, R.~Subramanian, Xia Tan, Binh
  Tang, Ross Taylor, Adina Williams, Jian~Xiang Kuan, Puxin Xu, Zhengxu Yan,
  Iliyan Zarov, Yuchen Zhang, Angela Fan, Melanie Kambadur, Sharan Narang,
  Aurelien Rodriguez, Robert Stojnic, Sergey Edunov, and Thomas Scialom.
  2023{\natexlab{b}}.
\newblock Llama 2: Open foundation and fine-tuned chat models.
\newblock \emph{ArXiv}, abs/2307.09288.

\bibitem[{Vaswani et~al.(2017)Vaswani, Shazeer, Parmar, Uszkoreit, Jones,
  Gomez, Kaiser, and Polosukhin}]{attention17}
Ashish Vaswani, Noam Shazeer, Niki Parmar, Jakob Uszkoreit, Llion Jones,
  Aidan~N. Gomez, Lukasz Kaiser, and Illia Polosukhin. 2017.
\newblock Attention is all you need.
\newblock In \emph{Advances in Neural Information Processing Systems 30: Annual
  Conference on Neural Information Processing Systems 2017, December 4-9, 2017,
  Long Beach, CA, {USA}}, pages 5998--6008.

\bibitem[{Voorhees(1999)}]{mrr}
Ellen~M. Voorhees. 1999.
\newblock The {TREC-8} question answering track report.
\newblock In \emph{Proceedings of The Eighth Text REtrieval Conference, {TREC}
  1999, Gaithersburg, Maryland, USA, November 17-19, 1999}, volume 500-246 of
  \emph{{NIST} Special Publication}. National Institute of Standards and
  Technology {(NIST)}.

\bibitem[{Wang and Wan(2018)}]{re_wang_18}
Ke~Wang and Xiaojun Wan. 2018.
\newblock Sentiment analysis of peer review texts for scholarly papers.
\newblock \emph{The 41st International ACM SIGIR Conference on Research \&
  Development in Information Retrieval}.

\bibitem[{Wang et~al.(2020)Wang, Zeng, Huang, Knight, Ji, and
  Rajani}]{re_wang_20}
Qingyun Wang, Qi~Zeng, Lifu Huang, Kevin Knight, Heng Ji, and Nazneen Rajani.
  2020.
\newblock Reviewrobot: Explainable paper review generation based on knowledge
  synthesis.
\newblock \emph{ArXiv}, abs/2010.06119.

\bibitem[{Wang et~al.(2022)Wang, Tang, Xia, Gong, Chen, and
  Liu}]{Wang2022CollaborativeFW}
Wei Wang, Tao Tang, Feng Xia, Zhiguo Gong, Zhikui Chen, and Huan Liu. 2022.
\newblock Collaborative filtering with network representation learning for
  citation recommendation.
\newblock \emph{IEEE Transactions on Big Data}, 8:1233--1246.

\bibitem[{Wright and Augenstein(2021)}]{citebert}
Dustin Wright and Isabelle Augenstein. 2021.
\newblock Citeworth: Cite-worthiness detection for improved scientific document
  understanding.
\newblock \emph{ArXiv}, abs/2105.10912.

\bibitem[{Wu et~al.(2023)Wu, Zheng, Qiu, Wang, Gu, Shen, Qin, Zhu, Zhu, Liu,
  Xiong, and Chen}]{Wu2023ASO}
Likang Wu, Zhilan Zheng, Zhaopeng Qiu, Hao Wang, Hongchao Gu, Tingjia Shen,
  Chuan Qin, Chen Zhu, Hengshu Zhu, Qi~Liu, Hui Xiong, and Enhong Chen. 2023.
\newblock A survey on large language models for recommendation.
\newblock \emph{ArXiv}, abs/2305.19860.

\bibitem[{Xie et~al.(2021)Xie, Zhu, Huang, Du, and Nie}]{Xie2021GraphNC}
Qianqian Xie, Yutao Zhu, Jimin Huang, Pan Du, and Jianyun Nie. 2021.
\newblock Graph neural collaborative topic model for citation recommendation.
\newblock \emph{ACM Transactions on Information Systems (TOIS)}, 40:1 -- 30.

\bibitem[{Yang et~al.(2023)Yang, Jin, Tang, Han, Feng, Jiang, Yin, and
  Hu}]{Yang2023HarnessingTP}
Jingfeng Yang, Hongye Jin, Ruixiang Tang, Xiaotian Han, Qizhang Feng, Haoming
  Jiang, Bing Yin, and Xia Hu. 2023.
\newblock Harnessing the power of llms in practice: A survey on chatgpt and
  beyond.
\newblock \emph{ArXiv}, abs/2304.13712.

\bibitem[{Yasunaga et~al.(2022)Yasunaga, Leskovec, and Liang}]{linkbert}
Michihiro Yasunaga, Jure Leskovec, and Percy Liang. 2022.
\newblock Linkbert: Pretraining language models with document links.
\newblock \emph{ArXiv}, abs/2203.15827.

\bibitem[{Yin and Li(2017)}]{Yin2017PersonalizedCR}
Jun Yin and Xiaoming Li. 2017.
\newblock Personalized citation recommendation via convolutional neural
  networks.
\newblock In \emph{APWeb/WAIM}.

\bibitem[{Yuan et~al.(2021)Yuan, Liu, and Neubig}]{re_yuan_21}
Weizhe Yuan, Pengfei Liu, and Graham Neubig. 2021.
\newblock Can we automate scientific reviewing?
\newblock \emph{J. Artif. Intell. Res.}, 75:171--212.

\bibitem[{Zhang and Ma(2020)}]{DACR}
Yang Zhang and Qiang Ma. 2020.
\newblock Dual attention model for citation recommendation.
\newblock In \emph{COLING}.

\bibitem[{Zheng et~al.(2023)Zheng, Xia, Zou, Dong, Wang, Xue, Shen, Wang, Wang,
  Li, Su, Yang, and Tang}]{codegeex}
Qinkai Zheng, Xiao Xia, Xu~Zou, Yuxiao Dong, Shan Wang, Yufei Xue, Lei Shen,
  Zi-Yuan Wang, Andi Wang, Yang Li, Teng Su, Zhilin Yang, and Jie Tang. 2023.
\newblock Codegeex: A pre-trained model for code generation with multilingual
  benchmarking on humaneval-x.
\newblock \emph{Proceedings of the 29th ACM SIGKDD Conference on Knowledge
  Discovery and Data Mining}.

\bibitem[{Zhou et~al.(2023)Zhou, Li, Li, Yu, Liu, Wang, Zhang, Ji, Yan, He,
  Peng, Li, Wu, Liu, Xie, Xiong, Pei, Yu, University, University, University,
  University, University, of~California~at San~Diego, University, of~Chicago,
  and Research}]{Zhou2023ACS}
Ce~Zhou, Qian Li, Chen Li, Jun Yu, Yixin Liu, Guan Wang, Kaichao Zhang, Cheng
  Ji, Qi~Yan, Lifang He, Hao Peng, Jianxin Li, Jia Wu, Ziwei Liu, Pengtao Xie,
  Caiming Xiong, Jian Pei, Philip~S. Yu, Lichao Sun Michigan~State University,
  Beihang University, Lehigh University, Macquarie University,
  Nanyang~Technological University, University of~California~at San~Diego, Duke
  University, University of~Chicago, and Salesforce~AI Research. 2023.
\newblock A comprehensive survey on pretrained foundation models: A history
  from bert to chatgpt.
\newblock \emph{ArXiv}, abs/2302.09419.

\bibitem[{Zhu et~al.(2023)Zhu, Yuan, Wang, Liu, Liu, Deng, Dou, and rong
  Wen}]{Zhu2023LargeLM}
Yutao Zhu, Huaying Yuan, Shuting Wang, Jiongnan Liu, Wenhan Liu, Chenlong Deng,
  Zhicheng Dou, and Ji~rong Wen. 2023.
\newblock Large language models for information retrieval: A survey.
\newblock \emph{ArXiv}, abs/2308.07107.

\end{thebibliography}

\bibliographystylelanguageresource{lrec_natbib}

\end{document}